\documentclass[onecolumn,showpacs,preprintnumbers,amsmath,amssymb]{revtex4}

\usepackage{graphicx} 
\usepackage{rotating}
\usepackage{bm}
\usepackage{bbm}
\usepackage{ulem} 
\usepackage[usenames]{color}

\newcommand{\eec}{\end{center}}
\newcommand{\bec}{\begin{center}}
\newcommand{\lamm}{$\Lambda^*(1520)\;$}
\newcommand{\be}{\begin{eqnarray}} 
\newcommand{\ee}{\end{eqnarray}}
\newcommand{\non}{\nonumber\\}

\begin{document}


\title{Radiative decay of the $\Delta^*(1700)$}

\author{M. D\"oring} \email{doering@ific.uv.es} \affiliation{
  Departamento de F\'{\i}sica Te\'orica and IFIC, \\ Universidad de
  Valencia--CSIC, 46100 Burjassot (Valencia), Spain}

\begin{abstract}
Electromagnetic properties provide information about the structure of strongly interacting systems and allow for independent tests of hadronic models. The radiative decay of the $\Delta^*(1700)$ is studied, which appears dynamically generated in a coupled channel approach from the rescattering of the $(3/2^+)$ decuplet of baryons with the $(0^-)$ octet of pseudoscalar mesons. The radiative decay is predicted from the well-known couplings of the photon to the mesons and hadrons which constitute this resonance in the dynamical picture.
\end{abstract}

\maketitle

\section{Introduction}
\label{sec:intro2}
The unitary extensions of chiral perturbation theory $(U\chi PT)$ have brought new light to the meson-baryon interaction, showing that some well-known resonances qualify as being dynamically generated. In this picture the Bethe-Salpeter resummation of elementary interactions, derived from chiral Lagrangians, guarantees unitarity and leads at the same time to genuine non-perturbative phenomena such as poles of the scattering amplitude in the complex plane of the invariant scattering energy $\sqrt{s}$, which can be identified with resonances. Coupled channel dynamics plays an essential role in this scheme, with the chiral Lagrangians providing the corresponding transitions of the multiplets; even physically closed channels contribute as intermediate virtual states. 

After earlier studies in this direction explaining the $\Lambda(1405)$ and the $N^*(1535)$ as meson-baryon $(MB)$ quasibound states \cite{Kaiser:1995cy,Kaiser:1996js,Oset:1997it,Nacher:1999vg,Inoue:2001ip,Oller:2000fj} from the interaction of the meson octet of the pion (M) with the baryon octet of the nucleon (B), new efforts have been undertaken \cite{Kolomeitsev:2003kt, Sarkar:2004jh} to investigate the low lying $3/2^-$ baryonic resonances which decay in $s$-wave into $0^-$ mesons $(M)$ and $3/2^+$ baryons $(B^*)$ of the decuplet. The leading interaction of these hadrons, 
given by the isovector term from Ref. \cite{Jenkins:1991es}, is unitarized by the use of the Bethe-Salpeter equation (BSE) in the on-shell reduction scheme from \cite{Sarkar:2004jh} which allows for a factorization of vertices and the intermediate loop function, thus reducing the BSE to an algebraic matrix equation in coupled channels. The unitarized amplitude develops poles in different isospin and strangeness channels in the complex plane of $\sqrt{s}$, which have been identified with resonances from the PDB \cite{Yao:2006px} such as the $\Lambda^*(1520)$, $\Sigma^*(1670)$, $\Delta^*(1700)$, etc. \cite{Sarkar:2004jh}.

In isospin $3/2$, strangeness 0, which are the quantum numbers of the $\Delta^*(1700)$, the coupled channels in $s$-wave are given by $\Delta(1232)\pi$, $\Sigma^*(1385)K$, and $\Delta(1232)\eta$. The $\Delta^*(1700)$, together with a series of other production mechanisms, has been included in Ref. \cite{Doring:2005bx}  in the study of the $\gamma p \to \pi^0 \eta p$ and $\gamma p \to \pi^0 K^0 \Sigma^+$ photoproduction reactions, currently measured at ELSA/Bonn. In the detailed work of \cite{Doring:2005bx} the $\Delta^*(1700)$, together with its strong couplings to $\Delta(1232)\eta$ and $\Sigma^*(1385)K$,  turned out to provide the dominant contribution. The branching ratios into these two channels are predicted from the scheme of dynamical generation and differ from a simple $SU(3)$ extrapolation of the $\Delta\pi$ channel by up to a factor of 30 \cite{Doring:2006pt}. 

The predictions for both reactions are in good agreement with preliminary data \cite{nanovanstar}. Recently, new measurements at low photon energies have been published \cite{Nakabayashi:2006ut} which also agree well with \cite{Doring:2005bx}. This has motivated another study \cite{Doring:2006pt} of altogether nine additional pion- and photon-induced reactions. From considerations of quantum numbers and the experimentally established $s$-wave dominance of the $\Sigma^*$ production close to threshold, the $\Delta^*(1700)$ channel is expected to play a major, in some reactions dominant, role. Indeed, good global agreement has been found for the studied reactions that span nearly two orders of magnitude in their respective cross sections. 

Thus, evidence from quite different experiments has been accumulated that the strong $\Delta^*(1700)\to \Sigma^*K, \,\Delta\eta$ couplings, predicted by the coupled channel model, are realistic. This gives support to the scheme of dynamical generation of this resonance. However, in all the photon-induced reactions from \cite{Doring:2005bx,Doring:2006pt} the initial $\gamma p\to\Delta^*(1700)$ transition has been taken from the experimental \cite{Yao:2006px} helicity amplitudes $A_{1/2}$ and $A_{3/2}$ \cite{Nacher:1998hh}: In Ref. \cite{Nacher:2000eq} the electromagnetic form factors $G_1'$, $G_2'$, and $G_3'$, which appear in the scalar and vector part of the $\gamma p\to\Delta^*(1700)$ transition, have been expressed in terms of the experimentally known $A_{1/2}$ and $A_{3/2}$ \cite{Yao:2006px}; this provides the transition on which we rely in all the photoproduction reactions via $\Delta^*(1700)$ in \cite{Doring:2005bx,Doring:2006pt}. 

Such a semi-phenomenological ansatz is well justified: the photon coupling and the width of the $\Delta^*(1700)$ is taken from phenomenology, whereas the strong decays of the $\Delta^*(1700)$ into hadronic channels are predictions from the unitary coupled channel model; the strengths of these strong transitions are responsible for the good agreement with experiment found in \cite{Doring:2005bx,Doring:2006pt}.
It is, however, straightforward to improve at this point, and this is the aim of this study. 

Electromagnetic properties provide additional information about the structure of strongly interacting systems and allow for an independent test of hadronic models, in this case the hypothesis that the $\Delta^*(1700)$ is dynamically generated. 
A virtue of the present model is that one can make predictions for the radiative decay, or equivalently, the inverse process of photoproduction; the components of the $\Delta^*(1700)$ in the meson-baryon 
base are all what matters, together with the well-known coupling of
the photon to these constituents. A similar study has been carried out for the radiative decay of the $\Lambda^*(1520)$ in Ref. \cite{Doring:2006ub} that has been described recently as a dynamically generated resonance \cite{Roca:2006sz}. For the $\Lambda^*(1520)\to\Sigma^0\gamma$ decay, where the dominant channels add up, indeed good agreement with experiment has been found. The present study is carried out along the lines of \cite{Doring:2006ub} but several modifications will be necessary: the $(\pi N)$ channel in $d$-wave plays an important role and is implemented in the coupled channel scheme. Second, a fully gauge invariant phototransition amplitude for the $s$-wave channels is derived that includes also couplings of the photon to the $B^*$ baryons.

\section{The model for the radiative $\boldsymbol{\Delta^*(1700)}$ decay}
In Sec. \ref{sec:improve} the coupled channel model from Ref. \cite{Sarkar:2004jh} is revised and extended to the inclusion of the $\pi N$ channel in $d$-wave, $(\pi N)_d$. In Sec. \ref{sec:phototrans} the model for the phototransition amplitude $\Delta^*(1700)\to\gamma N$ is derived: The photon interacts with the dynamically generated resonance via a one-loop intermediate state that is given by all coupled channels which constitute the resonance.

\subsection{The $(\pi N)_d$ channel in the unitary coupled channel approach}
\label{sec:improve}
In the unitarized model from Ref. \cite{Sarkar:2004jh} the $\Delta^*(1700)$ resonance appears as a quasi-bound state in the coupled channels $\Delta\pi$, $\Sigma^*K$, and $\Delta\eta$. The attraction in these channels leads to the formation of a pole in the $D_{33}$ channel which has been identified with the $\Delta^*(1700)$ resonance. 
In Ref. \cite{Sarkar:2004jh} the chiral interaction from \cite{Jenkins:1991es} is adapted in a nonrelativistic reduction providing isovector $MB^*\to MB^*$ transitions where $M\,(B^*)$ stands for the octet of $0^-$ pseudoscalar mesons ($3/2^+$ decuplet baryons). This interaction is unitarized by the use of the Bethe-Salpeter equation (BSE)
\be
T=(1-VG)^{-1}V
\label{again_BSE}
\ee
which turns out to be a matrix equation in coupled channels, factorized to an algebraic equation according to the on-shell reduction scheme of \cite{Sarkar:2004jh,Oller:1998zr}. In the recent work of \cite{Doring:2006ub} this scheme is compared in detail to other possible treatments of the BSE. The function $G$ is a diagonal matrix with the $MB^*$ loop functions $G_{MB^*}$ of the channels $i$ which are regularized in dimensional regularization with one subtraction constant $\alpha$. As it will appear in a different context in the phototransition, the function $G_{MB^*}$ has been re-derived with the result
\be
G_{MB^*}&=&i\int\frac{d^4p}{(2\pi)^4}\;\frac{2M}{(p+q)^2-M^2}\,\frac{1}{p^2-m^2}\non
&=&-2M\,\lim_{d\to 4}\left[
\int\frac{d^d\ell_E}{(2\pi)^d}\int\limits_0^1 dx\;\frac{1}{\left(
\ell_E^2+xM^2+(x-1)(xq^2-m^2)
\right)^2}\right]
\non&=&
\frac{2M}{(4\pi)^2}\bigg[\alpha+\log\frac{m^2}{\mu^2}+\frac{M^2-m^2+s}{2s}\;\log\frac{M^2}{m^2} +\frac{Q(\sqrt{s})}{\sqrt{s}}\;f_1(\sqrt{s})
\bigg]
\label{normal_MB}
\ee
where $m$ (M) is the meson (decuplet baryon) mass, $q^2\equiv s$ is the invariant scattering energy, and
\be
\alpha(\mu)=\gamma-\frac{2}{\epsilon}-\log{(4\pi)}-2,
\label{infi}
\ee
with the Eucledian integration over $\ell_E$ and $\epsilon=4-d$. The c.m. energy $\sqrt{s}$ in Eq. (\ref{normal_MB}) and all the following expressions of this study has to be taken at the physical sheet, i.e., $\sqrt{s}\to\sqrt{s}+i\epsilon$. The only exception is the pole search in the second Riemann sheet discussed at the end of this subsection.
Values of the regularization scale of $\mu=700$ MeV and $\alpha=-2$ are natural, as argued in \cite{Sarkar:2004jh}. The c.m. momentum function $Q$ and $f_1$ in Eq. (\ref{normal_MB}) are given by
\be
Q(\sqrt{s})&=&\frac{\sqrt{\left(s-(M+m)^2\right)\left(s-(M-m)^2\right)}}{2\sqrt{s}},\non
f_1(\sqrt{s})&=&\log\left(s-(M^2-m^2)+2\sqrt{s}\,Q(\sqrt{s})\right)+\log\left(s+(M^2-m^2)+2\sqrt{s}\,Q(\sqrt{s})\right)\non
&-&\log\left(-s+(M^2-m^2)+2\sqrt{s}\,Q(\sqrt{s})\right)-\log\left(-s-(M^2-m^2)+2\sqrt{s}\,Q(\sqrt{s})\right).
\label{qf1}
\ee
The loop function from Eq. (\ref{normal_MB}) has a real part, which implies a major difference of the present approach compared to unitarizations with the $K$-matrix. The real parts of the $G_i$, together with the attractive kernel $V$ in the isospin $3/2$, strangeness 0 channel, provide enough strength for the formation of a pole in the complex plane of the invariant scattering energy $\sqrt{s}$ which is identified with the $\Delta^*(1700)$.

However, additional channels will also couple to the dynamically generated resonance, changing in general its position and branching ratios, as these new channels can rescatter as well. In this study, the $(\pi N)_d$ channel is included in the analysis, because this is the lightest channel that can couple to the $\Delta^*(1700)$ and precise information of the $\pi N\to\pi N$ transition in the $D_{33}$ channel exists from the partial wave analysis (PWA) of Ref. \cite{Arndt:2006bf}. The $(\rho N)_s$ channel has been found important \cite{Yao:2006px,Manley:1984jz}, but for the radiative decay its influence is expected to be moderate as discussed below.

In order to include the $(\pi N)_d$ channel in the coupled channel model, one has to determine the $(\pi N)_d\to (MB^*)_s$ transitions, where $MB^*$ stands for the channels $\Delta\pi$, $\Sigma^*K$, and $\Delta\eta$ from \cite{Sarkar:2004jh}. There is no experimental information on these transitions. From the theoretical side, there is no information either due to the large number of low energy constants in the $d$-wave to $s$-wave transition. Thus, the coupling strengths have to be introduced as free parameters, called $\beta$. For the inclusion of $d$-wave potentials, we follow the lines of Ref. \cite{Sarkar:2005ap,Roca:2006sz} where it has been shown that the $d$-wave transitions can be factorized on-shell in the same way as the $s$-wave transitions; as a consequence, the meson-baryon $d$-wave loop function is the same as the $s$-wave loop function from Eq. (\ref{normal_MB}). We also allow for a $d$-wave to $d$-wave transition of the $(\pi N)_d$ channel.

With the channel ordering $i=1\cdots 4$ for $\Delta\pi$, $\Sigma^* K$, $\Delta\eta$, $(\pi N)_d$ the interaction kernel is given by
\be
V=\left(
\begin{array}{llll}
-\frac{2}{4\,f_\pi^2}\,\left(k^0+k'^0\right)&
-\frac{\sqrt{\frac{5}{2}}}{4\,f_\pi^2}\,\left(k^0+k'^0\right)&
0&
Q_{\pi N}^2\,r\,\beta_{(\pi N)_d\to\Delta\pi}\\
&-\frac{-1}{4\,f_\pi^2}\,\left(k^0+k'^0\right)&
-\frac{\frac{3}{\sqrt{2}}}{4\,f_\pi^2}\,\left(k^0+k'^0\right)&
Q_{\pi N}^2\,r\,\beta_{(\pi N)_d\to\Sigma^*K}\\
&&0&Q_{\pi N}^2\,r\,\beta_{(\pi N)_d\to\Delta\eta}\\
&&&Q_{\pi N}^4\,r^2\,\beta_{(\pi N)_d\to (\pi N)_d}\,\\
\end{array}
\right)
\label{newkern}
\ee
where we have multiplied some elements with $r=1/(4\cdot 93^2\cdot 1700)$ MeV$^{-3}$ = $1.7\cdot 10^{-8}$ MeV$^{-3}$ in order to obtain dimensionless transition strengths $\beta$ of the order of one. In Eq. (\ref{newkern}) one can recognize the chiral isovector transitions with $(k^0+k'^0)$ from Ref. \cite{Sarkar:2004jh} where $k^0=(s-M^2+m^2)/(2\sqrt{s})$ is the meson energy and $f_\pi=93$ MeV. 
In Eq. (\ref{newkern}), $Q_{\pi N}$ is the on-shell c.m. momentum of the $\pi N$ system and the $\beta$ are
the $s$-wave to $d$-wave transition strengths. See also Refs. \cite{Roca:2006sz, Sarkar:2005ap} where the analytic form of the transitions $V_{i4}$ ($i=1,\cdots,4$) has been derived.

Although a natural value for the subtraction constants is given by $\alpha=-2$ \cite{Sarkar:2004jh}, it is a common procedure \cite{Roca:2006sz,Inoue:2001ip} to absorb higher order effects in small variations around this value. Thus, as these higher order effects are undoubtly present, we allow for variations of the $\alpha$ of the four channels. Together with the transition strengths $\beta$, the set of free parameters is fitted to the single-energy-bin solution of the PWA of Ref. \cite{Arndt:2006bf}. Note that there is a conversion factor according to 
\be
\tilde{T}_{ij}(\sqrt{s})=-\sqrt{\frac{M_i Q_i}{4\pi\sqrt{s}}}\sqrt{\frac{M_j Q_j}{4\pi\sqrt{s}}}\;T_{ij}(\sqrt{s})
\label{conversion}
\ee
in order to express the solution $T$ of the Bethe-Salpeter equation (\ref{again_BSE}) in terms of the dimensionless amplitude $\tilde{T}(\sqrt{s})$ from \cite{Arndt:2006bf}. In Eq. (\ref{conversion}), $M_i$ ($Q_i$) is the baryon mass (c.m. momentum) of channel $i$; in the present case, $i=j=4$.

\begin{figure}
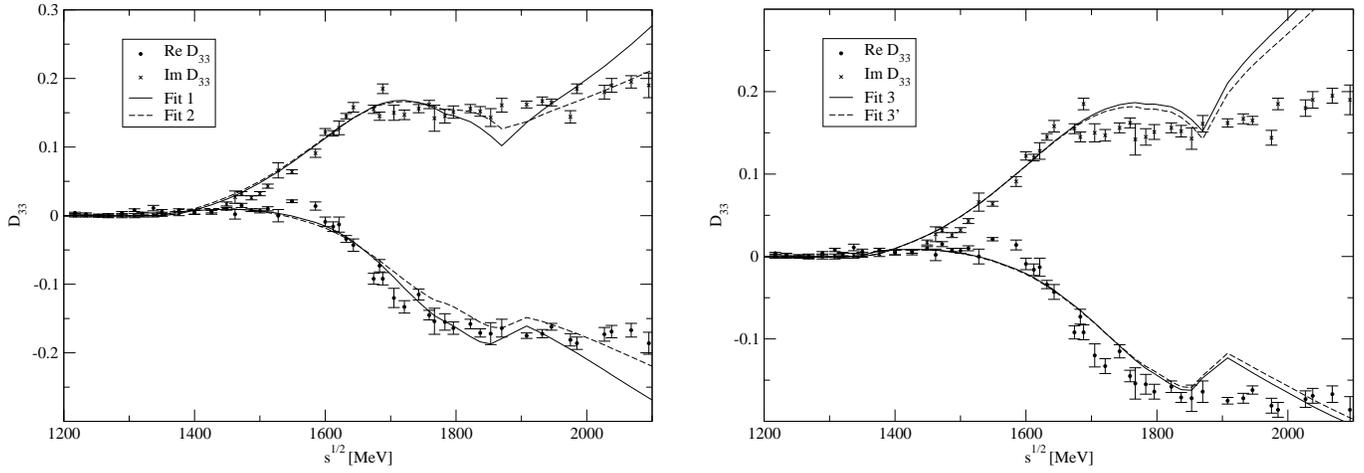

\bec
\includegraphics[width=0.48\textwidth]{fig_del_1_1.eps}
\hspace{0.5cm}
\includegraphics[width=0.48\textwidth]{fig_del_1_2.eps}
\caption{Fit results: Fit 1: All parameters free. Fit 2: without $\pi N\to\pi N$ $d$-wave transition kernel. Fit 3: As fit 2, but all subtraction constants for the $s$-wave loops chosen to be equal. Fit 3': As fit 3 but different minimum in $\chi^2$. The error bars show the single-energy-bin solution $\tilde{T}_{44}$ from Ref. \cite{Arndt:2006bf}.}
\label{fig:fitresults}
\eec
\end{figure}

For energies above 1.7 GeV a theoretical error of 0.08 has been added to the error bars from \cite{Arndt:2006bf} because additional channels such as $\rho N$ start to open and one can not expect good agreement much beyond the position of the $\Delta^*(1700)$. 
The resulting amplitudes of four different fits are plotted in Fig. \ref{fig:fitresults} with the parameter values displayed in Tab. \ref{tab:parms}.
\linespread{1.1}
\begin{table}
\caption{Parameter values of the fits to $(\pi N)_d$: the $\alpha_i$ are subtraction constants, the $\beta_i$ are $(\pi N)_d\to (B^*M)_s$ and $(\pi N)_d\to (\pi N)_d$ transition strengths. Note the sign changes of the $\beta_i$ between fit 3 and fit 3'.}
\begin{tabular*}{1\textwidth}{@{\extracolsep{\fill}}lllllrrrr}
\\ \hline \hline
&$\alpha_{\Delta\pi}$&$\alpha_{\Sigma^*K}$&$\alpha_{\Delta\eta}$&$\alpha_{(\pi N)_d}$&
$\beta_{(\pi N)_d\to\Delta\pi}$&$\beta_{(\pi N)_d\to\Sigma^*K}$&$\beta_{(\pi N)_d\to\Delta\eta}$& 
$\beta_{(\pi N)_d\to (\pi N)_d}$\vspace*{0.1cm}\\
\hline
Fit 1 &$-1.96$&$-1.28$&$-0.87$&$-1.00$&$2.19$&$-1.15$&$-0.07$&$63.5$\\
Fit 2 &$-1.25$&$-1.29$&$-0.66$&$-1.96$&$3.30$&$-0.37$&$0.31$&$0$\\
Fit 3 &$-1.21$&$-1.21$&$-1.21$&$-1.00$&$3.25$&$-0.85$&$-0.54$&$0$\\
Fit 3'&$-1.22$&$-1.22$&$-1.22$&$-1.00$&$-3.19$&$0.92$&$0.42$&$0$\\
\hline
\end{tabular*}
\label{tab:parms}
\end{table}
\linespread{1.0}
In fit 1, all parameters have been left free. The values for the $\beta$ are small: For $\sqrt{s}\sim M_{\Delta^*}$, they lead to values in the kernel (\ref{newkern}) around one order of magnitude smaller than the chiral interactions between channels 1 to 3; the size of $\beta_{(\pi N)_d\to (\pi N)_d}$ corresponds to a value  sometimes even two orders of magnitude smaller. These parameter values reflect the fact that the $\Delta^*(1700)$ couples only weakly to $\pi N$ and that the $\pi N$ interaction in $D_{33}$ is weak in general.

Thus, the $(\pi N)_d\to (\pi N)_d$ transition strength can be set to zero which is done for the fits 2, 3, and 3'.
One can even choose all subtraction constants of the $s$-wave channels to be equal, which is done in fit 3, and still obtain a sufficiently good result as Fig. \ref{fig:fitresults} shows. However, for fit three, there is another minimum in $\chi^2$, almost as good as the best one found. This fit is called fit 3'. As Tab. \ref{tab:parms} and Fig. \ref{fig:fitresults} show, one obtains an almost identical amplitude with a set of $\beta$'s with opposite sign as compared to fit 3 (see explanation below). 

For the different solutions 1 to 3', the coupling strengths of the resonance to the different channels can be obtained by expanding the amplitude around the resonance position in a Laurent series (see also Sec. \ref{sec:effphoto}). The residues give the coupling strengths which are uniquely determined up to a global sign which we fix by demanding the real part of the coupling to $\Delta\pi$ to be positive. In Tab. \ref{tab:couplings}
\linespread{1.1}
\begin{table}
\caption{Position $s_{\rm pole}^{1/2}$ and couplings of the $\Delta^*(1700)$. The values in brackets show the original results from \cite{Sarkar:2004jh} without the inclusion of the $(\pi N)_d$-channel. 
The PDB \cite{Yao:2006px} quotes a value of $s_{\rm pole}^{1/2}=(1620-1680)-i\,(160-240)$ MeV and couplings corresponding to $|g_{\Delta \pi}|=1.57\pm 0.3$, $|g_{(\pi N)_d}|=0.94\pm 0.2$. Note the sign change of $g_{(\pi N)_d}$ between fit 3 and fit 3'.}
\begin{tabular*}{1\textwidth}{@{\extracolsep{\fill}}lllllllll}
\\ \hline \hline
&$s_{\rm pole}^{1/2}$ [MeV]&$g_{\Delta\pi}$&$|g_{\Delta\pi}|$&$g_{\Sigma^*K}$&$|g_{\Sigma^*K}|$&$g_{\Delta\eta}$&$|g_{\Delta\eta}|$&$g_{(\pi N)_d}$ \vspace*{0.1cm}\\
&$(1827-i\,108)$&$(0.5+i\, 0.8)$&$(1.0)$&$(3.3+i\,0.7)$&$(3.4)$&$(1.7-i\,1.4)$&$(2.2)$&$(\;)$\\
\hline
Fit 1 &$1707-i\,160$&$1.09-i\,0.92$&$1.4$&$3.57+i\,1.91$&$4.0$&$-1.98-i\,1.68$&$2.6$&$-0.84-i\,0.05$\\
Fit 2 &$1692-i\,166$&$0.62-i\,1.03$&$1.2$&$3.44+i\,2.28$&$4.1$&$-1.89-i\,1.78$&$2.6$&$-0.77+i\,0.00$\\
Fit 3 &$1697-i\,214$&$0.68-i\,1.07$&$1.3$&$3.01+i\,1.95$&$3.6$&$-2.27-i\,1.89$&$3.0$&$-0.89+i\,0.15$\\
Fit 3'&$1698-i\,216$&$0.68-i\,1.07$&$1.3$&$3.02+i\,1.95$&$3.6$&$-2.27-i\,1.89$&$3.0$&$+0.92-i\,0.12$\\
\hline
\end{tabular*}
\label{tab:couplings}
\end{table}
\linespread{1.0}
the resulting couplings are displayed. The values in brackets quote the values of the original model from Ref. \cite{Sarkar:2004jh} without the inclusion of $(\pi N)_d$. Compared to these values, all couplings increase slightly in strength and have different phases. The largest change is observed for $|g_{\Delta\pi}|$ which has increased by 40 \% for fit 1; the new value is well inside the range quoted by the PDB \cite{Yao:2006px} that corresponds to $|g_{\Delta \pi}|=1.57\pm 0.3$. Also, the $d$-wave coupling to $\pi N$ coincides well with $|g_{(\pi N)_d}|=0.94\pm 0.2$ \cite{Yao:2006px}. The main properties of the resonance are conserved; in particular, the absolute values $|g|$ do not change much. This is relevant with respect to previous studies \cite{Doring:2005bx,Doring:2006pt} where the model from \cite{Sarkar:2004jh} has been used for the couplings of the $\Delta^*(1700)$ to $\Delta\eta$ and $\Sigma^* K$. 

Some explicit corrections to the results from \cite{Doring:2005bx,Doring:2006pt} are given, based on the values from fit 1 which is the preferred one as discussed below. 
The cross sections of the pion induced processes from \cite{Doring:2006pt} with the $\Sigma^*K$ final state will change by a factor $(|-0.84-i\,0.05|/0.94)^2(4.0/3.4)^2=1.1$, i.e., they stay practically the same. In the pion induced reaction with the $\Delta\eta$ final state the factor is again 1.1. Anticipating the result for fit 1 from Tab. \ref{tab:raddecay}, that the radiative coupling is close to the experimental one (which is the one used in  \cite{Doring:2005bx,Doring:2006pt}), the cross section of the reaction $\gamma p\to K^0\pi^+\Lambda$ increases by a factor of $(4.0/3.4)^2=1.4$ which improves the agreement with data, see Fig. 5 of \cite{Doring:2006pt} and also the recent measurements in Ref. \cite{thesis_wieland}. For $\gamma p\to\pi^0 K^0\Sigma^+$, which is also studied in \cite{Doring:2005bx,Doring:2006pt}, and where the photoproduction via the $\Delta^*(1700)$ dominates \cite{Doring:2005bx}, the factor is also close to 1.4. For the cross section of the $\gamma p\to\pi^0\eta p$ reaction, the transition via the $\Delta^*(1700)$ is only one of the reactions, and one has to re-evaluate the coherent sum of all processes of the model of \cite{Doring:2005bx}, with the new values from Tab. \ref{tab:couplings}. The numerical results for this reaction \cite{Doring:2005bx} have been updated in Fig. 5 of \cite{Doring:2006pt}. With the values of fit 1 from Tab. \ref{tab:couplings}, the cross section stays practically the same as in Fig. 5 of \cite{Doring:2006pt} (10 \% decrease).

Interestingly, the sign of the coupling to $(\pi N)_d$ is reversed in fit 3' compared to fit 3. This behavior has been noted before for the parameter values in Tab. \ref{tab:parms}. The reason for the difference between fit 3 and 3' is that the coupling of the $\pi N$ channel to the $\Delta^*(1700)$ is small: Although $(\pi N)_d$ is contained to all orders in the rescattering scheme, higher orders are smaller. Then, the (by far) dominant order in the fit to $(\pi N)_d\to(\pi N)_d$ is $g_{(\pi N)_d}^2$ and the relative sign to the other couplings $g_{\Delta\pi}$, $g_{\Sigma^*K}$, and $g_{\Delta\eta}$ is difficult to fix. As the influence of the $(\pi N)_d$ channel in the rescattering scheme is moderate, the main differences to the original model are caused by other parameters:  the important parameters, which have fine-tuned the original model from \cite{Sarkar:2004jh}, are the subtraction constants $\alpha_{\Delta\pi},\,\alpha_{\Sigma^*K},$ and $\alpha_{\Delta\eta}$. Their variation brings the pole down from $s_{\rm pole}^{1/2}=1827-i\,108$ MeV \cite{Sarkar:2004jh} to the values quoted in Tab. \ref{tab:couplings}.

In Sec. \ref{numresss} results for the radiative decay for all four fits are given, in order to obtain an idea of the systematic theoretical uncertainties. The fit 1 is preferred, though, because in the reduction to less free parameters, as it is the case for the fits 2, 3, and 3', the remaining free parameters have to absorb effects such as direct $(\pi N)_d\to(\pi N)_d$ transitions; results might become distorted, despite the fact that the fits appear to be good in Fig. \ref{fig:fitresults}. The larger space of free parameters also helps to fix the ambiguities found in fit 3 and 3': For fit 1 and 2, no alternative solutions with a reversed sign for $g_{(\pi N)_d}$ have been found, and, thus, the sign is fixed.

I have also performed a search for poles in the second Riemann sheet. For the different fits, the pole positions are given in Tab. \ref{tab:couplings}. The positions are in agreement with the values given by the PDB \cite{Yao:2006px}. Note that one of the virtues of the coupled channel analysis is that a separation of background and resonance part of the amplitude is not necessary; thus, the position of the pole does not suffer from ambiguities from this separation process required by other analyses.

Nevertheless, some theoretical uncertainties are present in the model from the omission of other channels such as $\rho N$ in $s$-wave or even $\rho N$, $\Delta\pi$ and $K\Sigma$ in $d$-wave, all of which have been reported in the PDB \cite{Yao:2006px}. Although the $\rho N$ channel is closed at the position of the $\Delta^*(1700)$, it contributes through the real part of the $\rho N$ loop function in the rescattering scheme, and through the finite width of the $\rho$ even to the imaginary part. 

However, in the calculation of the radiative decay, which is the aim of this study, no large contributions are expected from these heavy channels. In the study of the radiative decay of the $\Lambda^*(1520)$ \cite{Doring:2006ub} we have seen that contributions to the radiative decay width from the heavy channels are systematically suppressed, as also discussed in Sec. \ref{numresss}. 

\subsection{The phototransition amplitude}
\label{sec:phototrans}
\begin{figure}
\bec
\includegraphics[width=.8\textwidth]{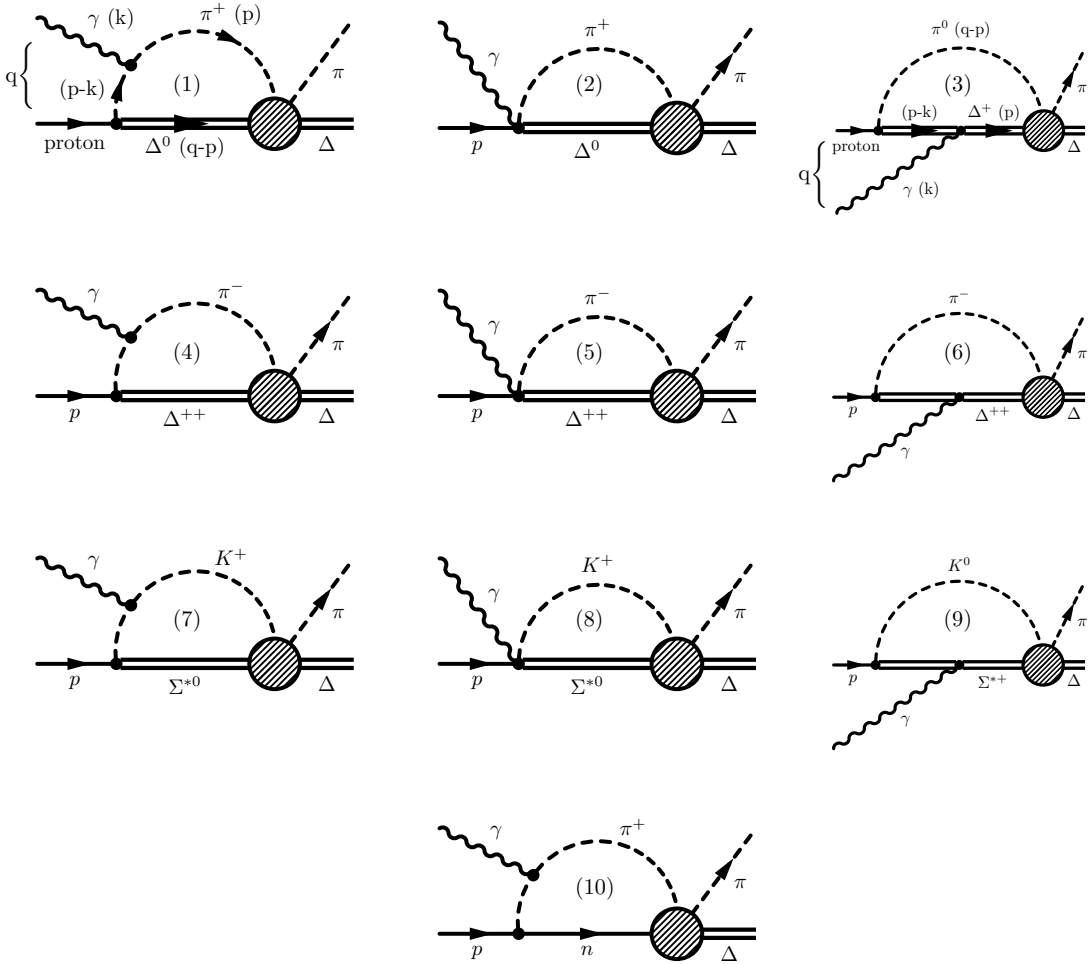}
\eec
\caption{Mechanisms for the $\Delta^*(1700)$ decay in $s$- and $d$-wave loops. The shaded circles represent the $\Delta^*(1700)$ in the transition $B^*M\to\Delta\pi$ or $\pi^+n\to\Delta\pi$. The diagrams in the left column are referred to as ''meson pole loops''. Diagrams (2), (5), and (8) have a $\gamma BMB^*$ transition and are referred to as ''Kroll-Ruderman loops'' (diagrams (3), (6), (9): ''baryon pole loops''). Diagram (10) shows the process with a $d$-wave coupling of the $(\pi N)_d$ intermediate state to the $\Delta^*(1700)$.}
\label{fig:delstardec}
\end{figure}

The only known radiative decay of the $\Delta^*(1700)$, which is also the one of relevance in Refs. \cite{Doring:2005bx,Doring:2006pt}, is into $\gamma N$ and we concentrate on this channel. The coupled channel model for the $\Delta^*(1700)$ has the virtue that the radiative decay can be calculated in a parameter-free and well-known way through the coupling of the photon to the particles which constitute the resonance.
The dominant photon couplings to the coupled channels $\Delta\pi$, $\Sigma^*K$, $\Delta\eta$, and $(\pi N)_d$ are displayed in Fig. \ref{fig:delstardec} .
We have chosen here the $I_3=+1/2$ or charge $C=+1$ state of the $\Delta^*(1700)$. The set of diagrams (1) to (9) is gauge invariant and finite as shown in Sec. \ref{sec:gi}.

There is also a loop diagram where the photon couples to the $\Delta^+$ in an intermediate $\Delta^+\eta$ state. This contribution is doubly suppressed: First, the diagrams with $\gamma B^*B^*$ couplings are much smaller than the other ones (see Sec. \ref{sec:gi}), and, second, the $\Delta\eta$ state is heavy which renders this contribution even smaller.

 Apart from the photon coupling to the $\Delta\pi$ and $\Sigma^*K$ channels, Fig. \ref{fig:delstardec} shows also the coupling to the $\pi N$ $d$-wave channel in diagram (10). Note that there is no Kroll-Ruderman term (which has $\pi N$ in $s$-wave in the $\gamma BMB$-vertex) as the vertex on the right hand side of the $\pi N$ loop is in $d$-wave, and the term vanishes in the integration over the loop momentum. The photon coupling to the $d$-wave loop is evaluated in Sec. \ref{sec:dwavepin}.

As for additional photon couplings, as e.g. to the external proton, to vertices of the rescattering, or to intermediate meson-baryon loops, apart from the ones considered, these processes are present \cite{Borasoy:2005zg} in general but negligible as discussed at the end of Sec. \ref{sec:gi}. 

The $MBB^*$ vertices appearing in Fig. \ref{fig:delstardec} are provided by the Lagrangian from Ref. \cite{Butler:1992pn}, with the relevant parts for the present reaction given by
\be
{\cal L}&=&{\cal C}\left(\overline{T}_\mu A^{\mu}B+{\overline B}A_\mu T^\mu\right)\non
&=&
{\cal C}\left(\sum_{a,b,c,d,e}^{1,\cdots,3}\epsilon_{abc}\;\overline{T}^{ade} \;\overline{u}_\mu\;A_{d}^{b,\mu} \;B^c_e
+\sum_{a,b,c,d,e}^{1,\cdots,3}\epsilon^{abc}\;{\overline B}^e_c \;A^{d}_{b,\mu} \;T_{ade}\;u^\mu\right)\non
\label{precise_LL}
\ee
where $B$ is the standard $3\times 3$ matrix of the $1/2^+$ baryon fields, $T$ is the field of the $3/2^+$ decuplet baryons with the definitions and  phase conventions of \cite{Butler:1992pn}, and $A^\mu$ is proportional to the axial current \cite{Butler:1992pn}. In the present situation, $A^\mu$ is expanded up to one meson field. From minimal coupling of the photon, the $\gamma BMB^*$ Kroll-Ruderman contact term is straightforward deduced from Eq. (\ref{precise_LL}). The constant ${\cal C}$ is fitted to phenomenologically known branching ratios. Explicit expressions for the required Feynman rules can be found, e.g., in \cite{Doring:2005bx,Doring:2006pt,Doring:2006ub}.

Let us start with the evaluation of diagram (1) in Fig. \ref{fig:delstardec}. The amplitude is given by 
\be
(-it)_{\pi^+\Delta^0}&=&-\,\frac{f_{\Delta\pi N}^*}{m}\sqrt{\frac{1}{3}}\,e\,S_\mu^\dagger\epsilon_\nu\;T_{\Delta\pi\to\Delta\pi}\non
&\times&
(-i)\,\int\frac{d^4p}{(2\pi)^4}\,\frac{2M}{(q-p)^2-M^2+i\epsilon}\,\frac{1}{p^2-m^2+i\epsilon}\,\frac{1}{(p-k)^2-m^2+i\epsilon}\;(p-k)^\mu(2p-k)^\nu
\label{dia1}
\ee
where $m(M)$ is the pion ($\Delta$) mass, $f_{\Delta\pi N}^*\equiv m_\pi/(\sqrt{2}\,f_\pi)\,{\cal C}_{\Delta\to N\pi}=2.13$ is the $\Delta\pi N$ coupling strength, $e^2=4\pi/137$ is the electric charge, $S_\mu^\dagger$ is the spin 1/2 $\to$ 3/2 transition operator which we approximate by $S_\mu^\dagger=(0,{\bf S}^\dagger)$, and $\epsilon_\nu$ is the polarization of the photon which in Coulomb gauge is given by $\epsilon_\nu=(0,\boldsymbol{\epsilon})$. The shaded circle in diagram (1) represents the $T$-matrix element $T_{\Delta\pi\to\Delta\pi}$ of the unitary coupled channel scheme in which the $\Delta^*(1700)$ appears dynamically generated.  In Sec. \ref{sec:effphoto} it will be matched to the results from Sec. \ref{sec:improve}.
Note the simplified structure of the $\Delta$-propagator in Eq. (\ref{dia1}). This is the same simplification as made in the model of dynamical generation \cite{Sarkar:2004jh} of the $\Delta^*(1700)$, see Eq. (\ref{normal_MB}).

In order to ensure gauge invariance, the loop in Eq. (\ref{dia1}) can be evaluated using a calculation technique from Refs. \cite{Marco:1999df,Roca:2006am}. In Sec. \ref{sec:gi} we will compare this scheme to a straightforward calculation of the loops of Fig. \ref{fig:delstardec}. The general structure of the loop function, or phototransition amplitude, is given by
\be
T^{\mu\nu}=a\,g^{\mu\nu}+b\,q^\mu q^\nu+c\,q^\mu k^\nu+d\,k^\mu q^\nu+e\, k^\mu k^\nu.
\label{generalt}
\ee
with the momenta $q$ and $k$ as defined in diagram (1). The terms with $c$ and $e$ do not contribute once contracted with $\epsilon^\nu$ according to Eq. (\ref{dia1}) and using the transversality of the photon, $\epsilon k=0$. The terms with $b$ and $d$ do not contribute as $\epsilon q=0$ in the c.m. frame where $|{\bf q}|=0$ and using the fact that $\epsilon^0=0$. Thus, the only term that will not vanish in Eq. (\ref{generalt}) is $(a\,g^{\mu\nu})$.

It can be shown that the sets of diagrams (1) to (6) and (7) to (9) of Fig. \ref{fig:delstardec} are gauge invariant.
Contracting $T^{\mu\nu}$ with the photon momentum $k^\nu$ and using the Ward identity $k_\nu\,T^{\mu\nu}\equiv 0$,
leads to the condition $a+d \,kq=0$. Note that diagram (2) of Fig. \ref{fig:delstardec} contributes only to the term with $a$, whereas diagram (1) contributes both to $a$ and $d$. However, evaluating $d$ (and from this, $a$ through the condition $a+d \,kq=0$) has the advantage that the loop integral is finite whereas both diagram (1) and (2) are logarithmically divergent. Using Feynman parameters and keeping only the terms proportional to $k^\mu q^\nu$ , the second line of Eq. (\ref{dia1}) becomes
\be
d\,k^\mu q^\nu =-\frac{4M\,k^\mu q^\nu}{(4\pi^2)}\int\limits_0^1 dx\int\limits_0^{1-x}dz\;\frac{x(z-1)}{x[(x-1)q^2+z(q^2-M_{\rm e}^2)+M^2]+(1-x)m^2},
\label{dterm}
\ee
where we have written the product $2qk=2q^0k^0=q^2-M_{\rm e}^2$ in the c.m. system where $|{\bf q}|=0$ and $M_{\rm e}$ is the mass of the external baryon, in this case a proton. Note that $k^0=|{\bf k}|=1/(2\sqrt{s})(s-M_{\rm e}^2)$ where $\sqrt{s}\equiv q^0$ which we will use several times in the following. 
From Eq. (\ref{dterm}) we calculate the term $a$ through the condition $a+d\,kq=0$,
\be
G_{{\rm g.i.}}^I&\equiv&a=-d\,kq\non
&=&\frac{2M}{(4\pi)^2}\bigg[\frac{M_{\rm e}^2-M^2+m^2}{2M_{\rm e}^2}+\frac{k\sqrt{s}\left[(M^2-m^2)^2-2M_{\rm e}^2M^2\right]-m^2 M_{\rm e}^4}{2\,M_{\rm e}^4\,s}\, \log\frac{M^2}{m^2}\non
&+&\frac{\left(M_{\rm e}^2-M^2+m^2 \right)\left(2M_{\rm e}^2 -s \right)}{4M_{\rm e}^3\,k\,\sqrt{s}}\;Q(M_{\rm e}) \;f_1(M_{\rm e})+\frac{M^2-M_{\rm e}^2-m^2+2k\,\sqrt{s}}{4\,k\,s}\;Q(\sqrt{s})\;f_1(\sqrt{s})\non
&+&\frac{m^2}{2\,k\,\sqrt{s}}
\bigg[{\rm Li}_2\left(\frac{-M^2+M_{\rm e}^2+m^2-2M_{\rm e}\,Q(M_{\rm e})}{2m^2}\right)+{\rm Li}_2\left(\frac{-M^2+M_{\rm e}^2+m^2+2M_{\rm e}\,Q(M_{\rm e})}{2m^2}\right)\non
&&-{\rm Li}_2\left(\frac{-M^2+s+m^2-2\sqrt{s}\,Q(\sqrt{s})}{2m^2}\right)
-{\rm Li}_2\left(\frac{-M^2+s+m^2+2\sqrt{s}\,Q(\sqrt{s})}{2m^2}\right)
\bigg]
\bigg]
\label{ggI}
\ee
which is gauge invariant by construction. 
In Eq. (\ref{ggI}), ${\rm Li}_2$ is the dilogarithm and $Q$ and $f_1$ are given in Eq. (\ref{qf1}).
Having calculated $a$ from $d$, the loop function $G_{{\rm g.i.}}^I$ corresponds to the meson pole diagram (1) from Fig. \ref{fig:delstardec} plus the Kroll-Ruderman term from diagram (2). 

Furthermore, the photon can also couple directly to the baryon as displayed in diagram (3). Only the non-magnetic part of the $\gamma B^*B^*$ coupling (convection term), given in \cite{Nacher:2000eq}, is considered. This term has the same structure and sign as the $\gamma MM$ vertex. For the magnetic part, see the discussion at the end of Sec. \ref{sec:gi}. Diagram (3) also contributes to the term $d\,k^\mu q^\nu$ in Eq. (\ref{generalt}). The contribution to $d$, let it be $d^{II}$, leads to an extra modification of the term $a$,
\be
G_{{\rm g.i.}}^{II}&\equiv& a^{II}=-d^{II}\,kq\non
&=&\frac{2M}{(4\pi)^2}\bigg[\frac{M_{\rm e}^2+M^2-m^2}{2M_{\rm e}^2}-\frac{k\sqrt{s}\,(M^2-m^2)^2-\,s\,M^2M_{\rm e}^2}{2\,M_{\rm e}^4\,s}\, \log\frac{M^2}{m^2}\non
&-&\frac{\left(m^2-M^2 \right)\left(2M_{\rm e}^2 -s \right)-s\,M_{\rm e}^2}{4M_{\rm e}^3\,k\,\sqrt{s}}\;Q(M_{\rm e}) \;f_1(M_{\rm e})-\frac{M^2-m^2+s}{4\,k\,s}\;Q(\sqrt{s})\;f_1(\sqrt{s})\non
&+&\frac{M^2}{2\,k\,\sqrt{s}}
\bigg[{\rm Li}_2\left(\frac{-m^2+M_{\rm e}^2+M^2-2M_{\rm e}\,Q(M_{\rm e})}{2M^2}\right)+{\rm Li}_2\left(\frac{-m^2+M_{\rm e}^2+M^2+2M_{\rm e}\,Q(M_{\rm e})}{2M^2}\right)\non
&&-{\rm Li}_2\left(\frac{-m^2+s+M^2-2\sqrt{s}\,Q(\sqrt{s})}{2M^2}\right)
-{\rm Li}_2\left(\frac{-m^2+s+M^2+2\sqrt{s}\,Q(\sqrt{s})}{2M^2}\right)
\bigg]
\bigg].
\label{ggII}
\ee
In order to determine the effective coupling of the photon to the $\Delta^*(1700)$ we construct isospin amplitudes from the diagrams given in Fig. \ref{fig:delstardec}. The isospin states of $\Delta\pi$ and $\Sigma^*K$ in $(I=3/2, I_3=1/2)$ are given by
\be
|\Delta\pi, I=3/2, I_3=1/2\rangle&=&\sqrt{\frac{2}{5}}\,|\Delta^{++}\pi^-\rangle+\sqrt{\frac{1}{15}}\,|\Delta^+\pi^0\rangle+\sqrt{\frac{8}{15}}\,|\Delta^0\pi^+\rangle,\non
|\Sigma^*K, I=3/2, I_3=1/2\rangle&=&\sqrt{\frac{1}{3}}\,|\Sigma^{*+}K^0\rangle+\sqrt{\frac{2}{3}}\,|\Sigma^{*0}K^+\rangle
\label{isoconventions}
\ee
with the phase convention $|\pi^+\rangle=-|1,1\rangle$. 

With the loop functions from Eqs. (\ref{ggI}) and (\ref{ggII}) and standard Feynman rules \cite{Doring:2005bx,Doring:2006pt,Doring:2006ub} we can calculate the isospin amplitudes for the sum of all $\Delta\pi$-loops and the $\Sigma^*K$-loops according to Eq. (\ref{isoconventions}) with the result
\be
(-i{\bf t}\cdot\boldsymbol{\epsilon})^{(I=3/2, I_3=1/2)}_{\gamma p\to\Delta\pi\to\Delta\pi}&=&\frac{\sqrt{10}}{3}\,e\,{\bf S}^\dagger\cdot \boldsymbol{\epsilon}\,\frac{f^*_{\Delta\pi N}}{m_\pi}\,\left(G_{{\rm g.i.}}^{I}+G_{{\rm g.i.}}^{II}\right)_{|m=m_\pi,\,M=M_\Delta, \,M_{\rm e}=M_N}\,T_{\Delta\pi\to\Delta\pi},\non
(-i{\bf t}\cdot\boldsymbol{\epsilon})^{(I=3/2, I_3=1/2)}_{\gamma p\to\Sigma^*K\to\Delta\pi}&=&\frac{1}{3\sqrt{2}}\,e\,{\bf S}^\dagger\cdot \boldsymbol{\epsilon}\,\frac{{\cal C}_{\Sigma^*\to N\bar{K}}}{f_\pi}\,\left(G_{{\rm g.i.}}^{I}+G_{{\rm g.i.}}^{II}\right)_{|m=m_K,\,M=M_{\Sigma^*}, \,M_{\rm e}=M_N}\,T_{\Sigma^*K\to\Delta\pi}
\label{firsttsdelstar}
\ee
where we have indicated which masses $m,\,M,\,M_{\rm e}$ have to be used in the loop functions.
In Eq. (\ref{firsttsdelstar}), 
\be
{\cal C}_{\Sigma^*\to N\bar{K}}=1.508\simeq \frac{6(D+F)}{5}.
\label{explicitbreaking}
\ee
The strength ${\cal C}_{\Sigma^*\to N\bar{K}}$ for the $\Sigma^*$ decay into the physically closed channel $N\bar{K}$ has been determined from from a $SU(6)$ quark model \cite{Oset:2000eg} in the same way as in Refs. \cite{Doring:2005bx,Doring:2006pt}: the SU(6) spin-flavor symmetry connects the $\pi NN$ coupling strength to the $\pi N\Delta$ strength, and then $SU(3)$ symmetry is used to connect the $\pi N\Delta$ transition with $\bar{K}N\Sigma^*$. The use of $SU(6)$ symmetry allows to express ${\cal C}_{\Sigma^*\to N\bar{K}}$ in terms of $D$ and $F$. 

In the decuplet, the $SU(3)$ symmetry is broken. This can be taken into account phenomenologically by allowing for different ${\cal C}$ in the Lagrangian (\ref{precise_LL}). For the open channels of the $\Sigma^*$ decay modes one obtains ${\cal C}_{\Sigma^*\to \Sigma\pi}=1.64$ and ${\cal C}_{\Sigma^*\to \Lambda\pi}=1.71$ from fitting to the partial decay widths into these channels. The constant ${\cal C}_{\Sigma^*\to N\bar{K}}=1.508$ from Eq. (\ref{explicitbreaking}) is close to these values (compare to ${\cal C}_{\Delta\to N\pi}=f^*_{\Delta\pi N}=2.13$).

\subsection{Gauge Invariance}
\label{sec:gi}
The construction of the gauge invariant amplitude in the last section can be compared to a straightforward calculation of the diagrams (1) to (9) in Fig. \ref{fig:delstardec}. In this section we show that both ways give identical results; at the end of this section we discuss further issues related to gauge invariance.

In Fig. \ref{fig:delstardec} there are three types of loops: the Kroll-Ruderman structure, the meson pole term, and the baryon pole term. All loops are logarithmically divergent and we calculate in dimensional regularization for the sake of conservation of gauge invariance.
The Kroll-Rudermann loop function is identical to the common meson-baryon loop function from Eq. (\ref{normal_MB}),
\be
G_{\gamma BMB^*}=G_{MB^*}.
\label{gkroll}
\ee
With the momenta assigned as in diagram (1) of Fig. \ref{fig:delstardec}, we define the meson pole loop function by
\be
&&(-i)\int\frac{d^4p}{(2\pi)^4}\,\frac{2M}{(q-p)^2-M^2+i\epsilon}\;\frac{1}{p^2-m^2+i\epsilon}\;\frac{1}{(p-k)^2-m^2+i\epsilon}\;(p-k)^\mu\,(2p-k)^\nu 
\non&\to&g^{\mu\nu}\;G_{\gamma MM}
\non&=&
\frac{g^{\mu\nu}M\left(2/\epsilon-\gamma+\log(4\pi)+\log\mu^2\right)}{(4\pi)^2}
-\frac{g^{\mu\nu}\,2M}{(4\pi)^2}\int\limits_0^1 dx\int\limits_0^{1-x} dz\;\log\left(x[(x-1)q^2+2zqk+M^2]+(1-x)m^2 \right)
\non
&=&g^{\mu\nu}\;\frac{2M}{(4\pi)^2}\;R(m,M)
\label{mmgamma}
\ee
with 
\be
R(m,M)&=&\frac{1}{2}\left(1-\alpha-\log\frac{m^2}{\mu^2}\right)
-\frac{(M^2-m^2)^2+sM_{\rm e}^2}{4\,s\,M_{\rm e}^2}\;\log\frac{M^2}{m^2}\non
&+&\frac{M_{\rm e}^2-M^2+m^2}{4M_{\rm e}\,k\,\sqrt{s}}\;Q(M_{\rm e})\;f_1(M_{\rm e})-\frac{s-M^2+m^2}{4\,k\,s}\;Q_(\sqrt{s})\;f_1(\sqrt{s})
\non&+&\frac{m^2}{2\,k\,\sqrt{s}}\bigg[{\rm Li}_2\left(\frac{-M^2+M_{\rm e}^2+m^2-2M_{\rm e}\,Q(M_{\rm e})}{2m^2}\right)+{\rm Li}_2\left(\frac{-M^2+M_{\rm e}^2+m^2+2M_{\rm e}\,Q(M_{\rm e})}{2m^2}\right)\non
&&-{\rm Li}_2\left(\frac{-M^2+s+m^2-2\sqrt{s}\,Q(\sqrt{s})}{2m^2}\right)
-{\rm Li}_2\left(\frac{-M^2+s+m^2+2\sqrt{s}\,Q(\sqrt{s})}{2m^2}\right)
\bigg]
\label{bstarbstargamma}
\ee
and $Q, \;f_1$ from Eq. (\ref{qf1}).
The arrow in Eq. (\ref{mmgamma}) indicates that we only keep the terms proportional to $g^{\mu\nu}$ because all other possible structures from Eq. (\ref{generalt}) do not contribute as commented following Eq. (\ref{generalt}). Similarly, and with the assignment of momenta as in diagram (3) of Fig. \ref{fig:delstardec}, the loop function where the photon couples directly to the baryon, is given by
\be
g^{\mu\nu}\;G_{\gamma B^*B^*}=g^{\mu\nu}\;\frac{2M}{(4\pi)^2}\;R(M,m).
\label{gbstarbstar}
\ee
Note that the convection part of the $\gamma B^*B^*$ coupling (the non-magnetic part) is of the same structure and sign as the $\gamma MM$ coupling \cite{Nacher:2000eq}.

The next step is to express the amplitudes of the diagrams in Fig. \ref{fig:delstardec} in terms of these three loop functions. Using standard Feynman rules \cite{Doring:2005bx,Doring:2006pt,Doring:2006ub}, we obtain for the diagrams (1) to (9)
\be
(-i{\bf t}\cdot\boldsymbol{\epsilon})_{(i)}=A_i\, e\, g_{B^* M B}\;{\bf S}^\dagger\cdot\boldsymbol{\epsilon}\, T_{B^*M\to\Delta\pi}
\label{allamp}
\ee
where $g_{B^* M B}=f^*_{\Delta\pi N}/m_\pi$ or $g_{B^* M B}={\cal C}_{\Sigma^*\to N\bar{K}}/(2f_\pi)$ from Eq. (\ref{explicitbreaking}) depending on whether $(B^*M)=(\Delta\pi)$ or $(\Sigma^*K)$. The coefficients $A_i$ for the diagrams (i) for $i=1$ to 9 are given in Tab. \ref{tab:ais}.
\begin{center}
\begin{table}
\caption{Coefficients $A_i$ for the diagrams (1) to (9) from Fig. \ref{fig:delstardec} with the amplitude given in Eq. (\ref{allamp}). 
The lower row shows how the infinities $2/\epsilon$ ($\epsilon=4-d$, see Eq. (\ref{infi}) for $\epsilon\to 0$) scale for each diagram. Once multiplied with the corresponding Clebsch-Gordan coefficients (CG) according to Eq. (\ref{isoconventions}), the sum over the infinities cancels, $\Sigma_i\;{\rm (CG)}_i\;r_i\left(\frac{2}{\epsilon}\right)=0$.}
\begin{tabular}{llllllllll}
\hline\hline
&(1)&(2)&(3)&(4)&(5)&(6)&(7)&(8)&(9)\vspace{-0.3cm}
\\ \\ \vspace{-0.3cm}
$A_i$\hspace*{0.4cm}&$\sqrt{\frac{1}{3}}G_{\gamma MM}$&$\sqrt{\frac{1}{3}}G_{\gamma BMB^*}$&$-\sqrt{\frac{2}{3}}G_{\gamma B^*B^*}$&$G_{\gamma MM}$&$G_{\gamma BMB^*}$&$2G_{\gamma B^*B^*}$&$\sqrt{\frac{1}{3}}G_{\gamma MM}$&$\sqrt{\frac{1}{3}}G_{\gamma BMB^*}$&$\sqrt{\frac{2}{3}}G_{\gamma B^*B^*}$
\\ \\
$r_i\left(\frac{2}{\epsilon}\right)\hspace*{0.4cm} $&$-\frac{1}{2}\left(\frac{2}{\epsilon}\right)$&$1\left(\frac{2}{\epsilon}\right)$&$\frac{1}{\sqrt{2}}\left(\frac{2}{\epsilon}\right)$&$-\frac{\sqrt{3}}{2}\left(\frac{2}{\epsilon}\right)$&$\sqrt{3}\left(\frac{2}{\epsilon}\right)$&$-\sqrt{3}\left(\frac{2}{\epsilon}\right)$&$-\frac{1}{2}\left(\frac{2}{\epsilon}\right)$&$1\left(\frac{2}{\epsilon}\right)$&$-\frac{1}{\sqrt{2}}\left(\frac{2}{\epsilon}\right)$
\\ \vspace{-0.3cm}\\ \hline\hline
\label{tab:ais}
\end{tabular}
\end{table}
\end{center}
The sum over all diagrams results in 
\be
(-i{\bf t}\cdot\boldsymbol{\epsilon})^{(I=3/2, I_3=1/2)}_{\gamma p\to\Delta\pi\to\Delta\pi}&=&\frac{\sqrt{10}}{3}\,e\,{\bf S}^\dagger\cdot \boldsymbol{\epsilon}\,\frac{f^*_{\Delta\pi N}}{m_\pi}\,\left(G_{\gamma MM}+G_{\gamma BMB^*}+G_{\gamma B^*B^*}\right)_{|m=m_\pi,\,M=M_\Delta, \,M_{\rm e}=M_N}\,T_{\Delta\pi\to\Delta\pi},\non
(-i{\bf t}\cdot\boldsymbol{\epsilon})^{(I=3/2, I_3=1/2)}_{\gamma p\to\Sigma^*K\to\Delta\pi}&=&\frac{1}{3\sqrt{2}}\,e\,{\bf S}^\dagger\cdot \boldsymbol{\epsilon}\,\frac{{\cal C}_{\Sigma^*\to N\bar{K}}}{f_\pi}\,\left(G_{\gamma MM}+G_{\gamma BMB^*}+G_{\gamma B^*B^*}\right)_{|m=m_K,\,M=M_{\Sigma^*}, \,M_{\rm e}=M_N}\,T_{\Sigma^*K\to\Delta\pi}.\non
\label{secdelstar}
\ee
The infinities of the nine diagrams, i.e., the terms with $2/\epsilon$ from the Eqs. (\ref{gkroll}, \ref{mmgamma}, \ref{gbstarbstar}) scale as shown in the lower row of Tab. \ref{tab:ais}. In order to construct the isospin states according to Eq. (\ref{isoconventions}) each infinity is multiplied with the corresponding Clebsch-Gordan coefficient; as a result, the sums over all infinities cancel for the $\Delta\pi$ loops and also for the $\Sigma^*K$ loops. Note that the phase convention $|\pi^+\rangle=-|1,1\rangle$ is crucial at this point. The Ward identity is working. In other words, gauge invariance renders the phototransition amplitude finite and leads to a parameter-free expression.
Comparing Eqs. (\ref{secdelstar}) and (\ref{firsttsdelstar}), obviously 
\be
G_{\gamma MM}+G_{\gamma BMB^*}+G_{\gamma B^*B^*}=G_{{\rm g.i.}}^{I}+G_{{\rm g.i.}}^{II}
\label{allgammas}
\ee
which can also be seen by comparing the explicit expressions given in Eqs. (\ref{ggI}, \ref{ggII}, \ref{gkroll}, \ref{mmgamma}, \ref{gbstarbstar}). In other words, the scheme from Eq. (\ref{generalt}), which allows for the construction of a gauge invariant amplitude through the condition $a+d\,kq=0$, leads to the same result as a straightforward calculation of the amplitude, in which the infinities cancel systematically. However, this is only the case if all contributions to $d$ are taken into account, in the present case from the meson pole term plus the baryon pole term ($G_{{\rm g.i.}}^{I}$ and $G_{{\rm g.i.}}^{II}$).

In the rest of this section, further issues of gauge invariance are discussed such as a comparison to cut-off schemes, the role of magnetic couplings, and gauge invariance in the context of the rescattering scheme.

In several recent studies \cite{Doring:2005bx,Doring:2006ub} the occurring photon loops have been regularized with a cut-off. As we have now a gauge invariant, parameter-free scheme at hand, we would like to compare both methods numerically. As a first test, the scheme has been implemented in the calculation of the radiative decay width of the $\Lambda^*(1520)$ from Ref. \cite{Doring:2006ub}. This means a gauge invariant evaluation of the $s$-wave loops from Fig. 2 of \cite{Doring:2006ub} formed by $\pi\Sigma^*$ and $K\Xi^*$, plus additional diagrams with $\gamma \Sigma^* \Sigma^*$ and $\gamma\Xi^*\Xi^*$ couplings in analogy to the diagrams in Fig. \ref{fig:delstardec}. 
In practice, the re-calculation only requires the replacement of the terms $\left(G_i+\frac{2}{3}\;\tilde{G}_i\right)$ from Eq. (41) of \cite{Doring:2006ub} by $\left(G_{{\rm g.i.}}^{I}+G_{{\rm g.i.}}^{II}\right)$ or $G_{\gamma MM}+G_{\gamma BMB^*}+G_{\gamma B^*B^*}$ from Eq. (\ref{allgammas}).
The final result from Ref. \cite{Doring:2006ub} for the radiative decay $\Lambda^*(1520)\to\gamma\Sigma^0$ changes from $\Gamma=60$ keV \cite{Doring:2006ub} to $\Gamma=61$ keV. Thus, the approximations made in \cite{Doring:2006ub} and the violation of gauge invariance are well under control. Note that the additional diagram with a $\gamma \Sigma^* \Sigma^*$ coupling cancels for the second radiative decay studied  in \cite{Doring:2006ub}, $\Lambda^*(1520)\to\gamma\Lambda$, in the same way as the $\pi\Sigma^*$ meson pole and Kroll-Ruderman terms. This is a consequence of gauge invariance but can be also seen directly by noting that the $\gamma MM$ interaction and the convection term of the $\gamma B^*B^*$ interaction have the same structure and sign \cite{Nacher:2000eq}. Thus, the $\Lambda^*(1520)\to\gamma\Lambda$ radiative width stays as small as already found in \cite{Doring:2006ub}.

In the cut-off scheme, the meson pole loop is defined as 
\be
\tilde{G}^{{\rm (cut)}}_i&=&i\;\int\frac{d^4 q}{\left(2\pi\right)^4}\;\frac{ {\bf q}^2-\left({\bf q}\cdot{\bf k}\right)^2/|{\bf k}|^2}{(q-k)^2-m_i^2+i\epsilon}\;\frac{1}{q^2-m_i^2+i\epsilon}\;
\frac{M_i}{E_i({\bf q})}\;
\frac{1}{P^0-q^0-E_i({\bf q})+i\epsilon},\non
&=&-\int\limits_0^\Lambda\frac{dq\;q^2}{\left(2\pi\right)^2}\int\limits_{-1}^1 dx\;\frac{ q^2(1-x^2)}{2\omega_i\omega_i'}\;\frac{1}{k+\omega_i+\omega_i'}\;\frac{1}{k-\omega_i-\omega_i'+i\epsilon}\;
\frac{M_i}{E_i(q)}
\non&\times&
\frac{1}{\sqrt{s}-\omega_i-E_i(q)+i\epsilon}\;\frac{1}{\sqrt{s}-k-\omega_i'-E_i( q)+i\epsilon}
\;
\left[\left(\omega_i+\omega_i'\right)^2+\left(\omega_i+\omega_i'\right)\left(E_i( q)-\sqrt{s}\right)+k\omega_i'\right].
\label{tildegexact}
\ee 
For consistency, the notation is as in \cite{Doring:2006ub}; $x$ is the cosinus of the angle between ${\bf q}$ and ${\bf k}$ with ${\bf k}$ the momentum of the real photon ($|{\bf k}|\equiv k$); $m_i$ is the meson mass, $P^0\equiv\sqrt{s}$, and $\omega_i, \,\omega'_i$ are the energies of the mesons at momentum $q$ and $q-k$, respectively; $E_i$ the energy of the baryon. Note that Eq. (\ref{tildegexact}) is slightly different from the corresponding expression in \cite{Doring:2006ub}, called $\tilde{G}$ there, as one kinematical approximation made in \cite{Doring:2006ub}, the angle average over the term $1-x^2$, is not sufficient in the present case, because the momenta of the decay products are higher. One obtains $\tilde{G}$ from \cite{Doring:2006ub} by substituting $(1-x^2)\to 1$ in Eq. (\ref{tildegexact}). Note that $\tilde{G}^{{\rm (cut)}}$ corresponds to $2/3$ of $\tilde{G}$.

The meson pole term $\tilde{G}^{{\rm (cut)}}$ is accompanied by the corresponding Kroll-Ruderman term with cut-off, called $G^{{\rm (cut)}}$, where explicit expressions can be found, e.g., in Eq. (22) of \cite{Doring:2006ub}. The cut-off can be determined by requiring the real part of the Kroll-Ruderman loop function to be equal in both dimensional regularization and cut-off scheme, at the energy of the resonance. For $\Delta\pi$ in the loop, this leads to a cut-off of $\Lambda=881$ MeV. 
\begin{figure}
\bec
\includegraphics[width=0.49\textwidth]{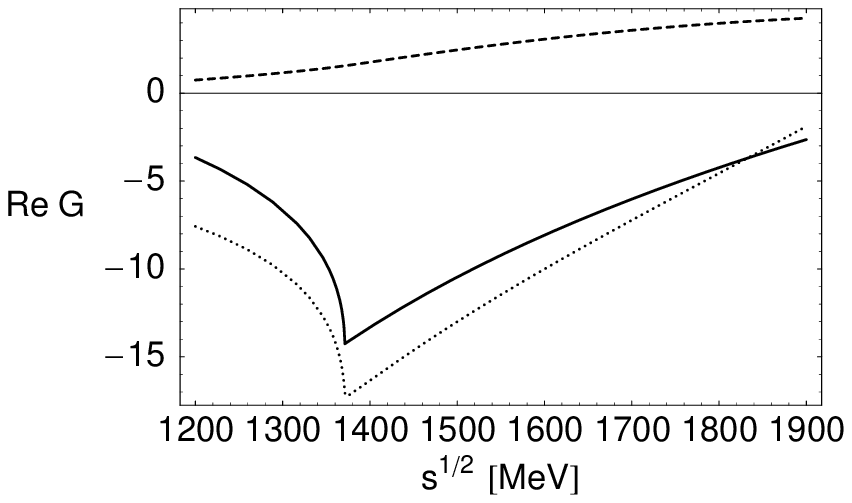}
\includegraphics[width=0.49\textwidth]{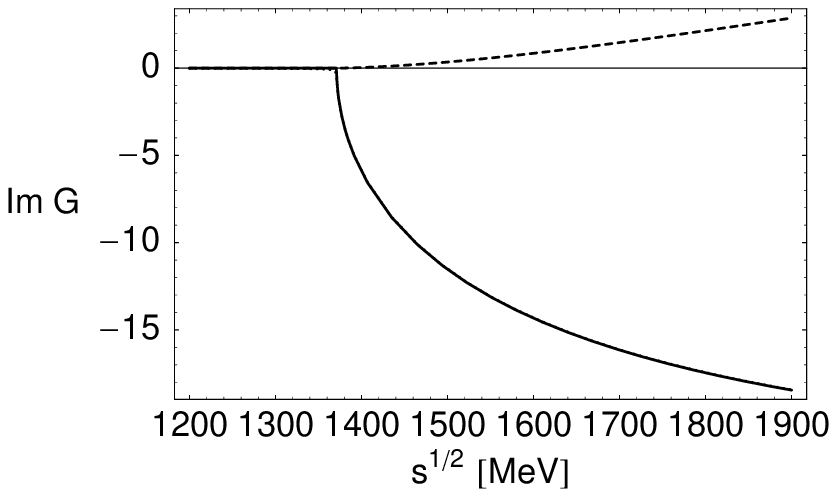}
\caption{Real and imaginary parts of $\Delta\pi$ loop functions. Solid line: gauge invariant $G_{{\rm g.i.}}^{I}$ (meson pole term plus Kroll-Ruderman term). Dashed line: gauge invariant $G_{{\rm g.i.}}^{II}$ (baryon pole term). Dotted line: meson pole plus Kroll-Ruderman term in a cut-off scheme ($\tilde{G}^{{\rm (cut)}}+G^{{\rm (cut)}}$) with $\Lambda=881$ MeV. The cut-off scheme and $G_{{\rm g.i.}}^{I}$ have identical imaginary parts}
\label{fig:props}
\eec
\end{figure}

In Fig. \ref{fig:props} the cut-off loops for $\Delta\pi$ ($\tilde{G}^{{\rm (cut)}}+G^{{\rm (cut)}}$), are shown as the dotted line.  The gauge invariant function $G_{{\rm g.i.}}^{I}$ from Eq. (\ref{ggI}) is plotted with the solid line. The imaginary parts of both results are identical as expected, but, more interestingly, at the energies of the $\Delta^*$ of around 1700 MeV, also the real parts coincide closely. The dashed line shows the gauge invariant function $G_{{\rm g.i.}}^{II}$ from Eq. (\ref{ggII}). This contribution comes from the baryon pole diagrams. As there are two baryon propagators, the diagram should be smaller which is indeed the case as Fig. \ref{fig:props} shows. However, results can be affected noticeable and one should include this term in general.

From the comparison in Fig. \ref{fig:props} we see that the cut-off scheme as it has been used in \cite{Doring:2006ub} (Kroll-Ruderman plus meson pole term) indeed takes into account the dominant contributions. The baryon pole term, which has not been considered in \cite{Doring:2006ub}, is small. If we would take this term into account in the cut-off scheme, the expression would be finite and in the limit $\Lambda\to \infty$, both cut-off scheme and gauge invariant scheme would give identical results. 

Finally, there are photon couplings to the external baryon of the rescattering scheme, to vertices of the rescattering scheme itself, and to intermediate loops of the rescattering scheme which all have been ignored in the present study. This is because with these couplings, the first loop has no photon attached any longer and is effectively suppressed because the integration over the momentum vanishes due to the presence of one $s$-wave and one $p$-wave vertex, or one $s$-wave and one $d$-wave vertex. A detailed discussion can be found in Ref. \cite{Doring:2005bx}. There is also the magnetic part of the $\gamma B^*B^*$ vertex which is proportional to ${\bf S}_\Delta\times {\bf k}$ \cite{Nacher:2000eq} with k the photon momentum. As evaluated in Eq. (11) of \cite{Doring:2005bx}, this contribution vanishes for large baryon masses and in practice can be neglected.

\subsection{Photon coupling to the $\pi N$ loop in $d$-wave}
\label{sec:dwavepin}
Diagram (10) of Fig. \ref{fig:delstardec} shows the phototransition amplitude via the $\pi N$ state in $d$-wave. 
The evaluation follows the same steps as in Sec. \ref{sec:phototrans}.
For $\pi N$ in $(I,I_3)=(3/2,1/2)$, which is the configuration chosen here, $|N\pi\rangle=-\sqrt{1/3}\,|n\pi^+\rangle+\sqrt{2/3}\,|p\pi^0\rangle$. 
Using  standard Feynman rules \cite{Doring:2006ub}, we obtain
\be
(-i{\bf t}\cdot\boldsymbol{\epsilon})^{(I=3/2, I_3=1/2)}_{\gamma p\to N\pi\to\Delta\pi}&=&-\,\frac{\sqrt{2}}{3}\,e\,{\bf S}^\dagger\cdot \boldsymbol{\epsilon}\,\frac{D+F}{f_\pi}\,\tilde{G}'_{N\pi}\,T_{(\pi N)_d\to\Delta\pi}.
\label{minitdwavedell}
\ee
The $d$-wave meson pole loop function $\tilde{G}'_{N\pi}$ has been calculated in Refs. \cite{Doring:2006ub} and \cite{Hyodo:2006uw} and we use the results from there. In the notation from \cite{Doring:2006ub},
\be
{\tilde G}'_{N\pi}&=&i\;\int\frac{d^4 q}{\left(2\pi\right)^4}\;\frac{{\bf q}^2}{(q-k)^2-m_\pi^2+i\epsilon}\;\frac{1}{q^2-m_\pi^2+i\epsilon}\;
\frac{1}{P^0-q^0-E_N({\bf q})+i\epsilon}\;\frac{M}{E_N({\bf q})}\left(\frac{{\bf q}^2}{Q^2_{\pi N}(M_{\Delta^*})}\right)
\label{gtildep}
\ee
where $Q_{\pi N}(M_{\Delta^*})$ is the on-shell three-momentum of pion and nucleon for the $\Delta^*(1700)$ decay at rest; $M$ is the nucleon mass and the other quantities are defined as in Eq. (\ref{tildegexact}). 

The development of a gauge invariant scheme as for the $s$-wave loops in Sec. \ref{sec:phototrans}, \ref{sec:gi} is beyond the scope of this work. Instead, we follow the lines of Ref. \cite{Doring:2006ub}. In the numerical comparison at the end of Sec. \ref{sec:gi} we have seen that the violation of gauge invariance from the cut-off scheme leads almost to the same results as the gauge invariant calculation, at least sufficiently beyond threshold; thus, the loop function in Eq. (\ref{gtildep}) is regularized with a cut-off.

The cut-off is determined from a comparison between the $d$-wave meson-baryon loop in dimensional regularization with on-shell factorization of the vertices (i.e., as it appears in the rescattering scheme of Sec. \ref{sec:improve}), and the $d$-wave loop with cut-off according to 
\be
Q_{\pi N}^4(\sqrt{s}) \;G_{\pi N}(\sqrt{s})= \int\limits_0^\Lambda\frac{dp\, p^2}{2\pi^2}\;\frac{p^4}{2\omega}\;\frac{M}{E(p)}\;\frac{1}{\sqrt{s}-\omega(p)-E(p)+i\epsilon}
\label{matchingd}
\ee
at the energy $\sqrt{s}$ of the real part of the resonance position given in Tab. \ref{tab:couplings}. In Eq. (\ref{matchingd}), $G_{\pi N}$ is the loop in dimensional regularization from Eq. (\ref{normal_MB}) with the subtraction constants $\alpha_{(\pi N)_d}$ from Tab. \ref{tab:parms}; $Q_{\pi N}^4$ is the on-shell c.m. momentum from the two $d$-wave vertices in on-shell factorization, $\omega$ ($E$) is the pion (nucleon) energy, and $M$ the nucleon mass. The cut-off determined in this way is used for the regularization of the meson pole term of Eq. (\ref{gtildep}).

\subsection{Effective photon coupling}
\label{sec:effphoto}
\begin{figure}
\centerline{\includegraphics[width=0.3\textwidth]{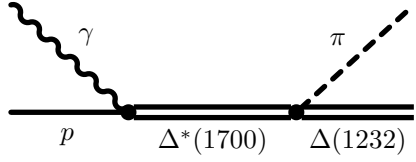}}
\caption{Effective resonance representation of the radiative decay.}
\label{fig:mech2}
\end{figure}

In the last sections the amplitudes for the process $\gamma p\stackrel{{\Delta^*(1700)}}{\longrightarrow} \Delta\pi$ have been determined and are written in terms of the $T^{(i1)}$, the unitary solution of the BSE (\ref{again_BSE}) for meson-baryon scattering with the transitions from channel $i$ ($\Delta\pi$, $\Sigma^*K$, $(\pi N)_d$) to the $\Delta\pi$ final state (channel no. 1). In order to determine the partial photon decay width of the $\Delta^*(1700)$, the $T^{(i1)}$ are expanded around the simple pole in the complex plane with the leading term of the Laurent series given by
\be
T^{(i1)}\simeq\frac{g_i g_{\Delta\pi}}{\sqrt{s}-M_{\Delta^*(1700)}}
\label{texpa}
\ee
where $g_i$ is the $\Delta^*(1700)$ coupling to channel $i$, the product $g_i g_{\Delta\pi}$ is provided by the residue and $M_{\Delta^*(1700)}$ is the complex pole position given in Tab. \ref{tab:couplings}.
With this replacement for the $T^{(i1)}$, the amplitudes from Eqs. (\ref{firsttsdelstar}, \ref{minitdwavedell}) can be matched to the resonant process shown in Fig. \ref{fig:mech2}, 
which is given by
\be
\left(-i {\bf t}\cdot\boldsymbol{\epsilon}\right)=\left(-ig_{\Delta\pi}\right)\;\frac{i}{\sqrt{s}-M_{\Delta^*}}
\;g_{\Delta^*\gamma p}\;{\bf S}^\dagger\cdot\boldsymbol{\epsilon}.
\label{treson}
\ee
This identification allows to write the effective $\Delta^*(1700)\gamma p$ coupling ($p$=proton), $g_{\Delta^*\gamma p}$, in terms of the one-loop photoproduction processes discussed in the last sections:
\be
g_{\Delta^*\gamma p}^{(\Delta\pi)}&=&e\;g_{\Delta\pi}\;\frac{\sqrt{10}}{3}\, \frac{f^*_{\Delta\pi N}}{m_\pi}\left(
G_{{\rm g.i.}}^{I}+G_{{\rm g.i.}}^{II}
\right)\non
g_{\Delta^*\gamma p}^{(\Sigma^*K)}&=&e\;g_{\Sigma^*K}\;\frac{1}{3\sqrt{2}}\,\frac{{\cal C}_{\Sigma^*\to N\bar{K}}}{f_\pi}
\left(
G_{{\rm g.i.}}^{I}+G_{{\rm g.i.}}^{II}
\right)
\ee
with the couplings $g$ of the $\Delta^*(1700)$ given in Tab. \ref{tab:couplings} and the $G$ from Eq. (\ref{allgammas}). 
In the same way, the $d$-wave amplitude in Eq. (\ref{minitdwavedell}) is matched, resulting in
\be
g_{\Delta^*\gamma p}^{(N\pi)}&=&-\;g_{(\pi N)_d}\,e\,\frac{\sqrt{2}}{3}\,\frac{D+F}{f_\pi}\;\tilde{G}'_{N\pi}.
\label{nocheing}
\ee
The effective photon coupling is given by the coherent sum of all processes from Fig. \ref{fig:delstardec},
\be
g_{\Delta^*\gamma p}=g_{\Delta^*\gamma p}^{(\Delta\pi)}+g_{\Delta^*\gamma p}^{(\Sigma^*K)}+g_{\Delta^*\gamma p}^{(N\pi)}.
\label{sumgg}
\ee

\section{Numerical results}
\label{numresss}
The results for the radiative decay width, given by
\be
\Gamma_{\Delta^*\to\gamma N}=\frac{k}{3\pi}\,\frac{M_p}{M_{\Delta^*}}\;|g_{\Delta^*\gamma p}|^2
\ee
with $g_{\Delta^*\gamma p}$ from Eq. (\ref{sumgg}) and $k$ the c.m. momentum of the photon, are summarized in Tab. \ref{tab:raddecay}.	
\linespread{1.1}
\begin{table}
\caption{Radiative decay width $\Gamma$ of the $\Delta^*(1700)$, to be compared with $\Gamma=\boldsymbol{570}\pm \boldsymbol{254}$ keV from the PDB \cite{Yao:2006px}. Also, the effective couplings $g_{\Delta^*\gamma p}^{(\cdots)}$ of the photon to the $\Delta^*(1700)$ via $\Delta\pi$, $\Sigma^*K$, and $(\pi N)_d$ loops are displayed, in order to show the interference pattern of the channels (multiplied with $10^3$).}
\begin{tabular*}{0.6\textwidth}{@{\extracolsep{\fill}}lrrrr}
\\ \hline \hline
&$g_{\Delta^*\gamma p}^{(\Delta\pi)}$&$g_{\Delta^*\gamma p}^{(\Sigma^*K)}$&$g_{\Delta^*\gamma p}^{(N\pi)}$&$\Gamma$ [keV] \vspace*{0.1cm}\\
\hline
Fit 1 &$-82-i\,71$&$-30-i\,16$&$-9+i\,36$&{\bf 602}\\
Fit 2 &$-85-i\,32$&$-28-i\,18$&$6+i\,34$&{\bf 403}\\
Fit 3 &$-89-i\,37$&$-25-i\,16$&$-1+i\,40$&{\bf 459}\\
Fit 3'&$-89-i\,37$&$-25-i\,16$&$3-i\,41$&{\bf 730}\\
\hline
\end{tabular*}
\label{tab:raddecay}
\end{table}
\linespread{1.0}
For the experimental value of $\Gamma=570\pm 254$ keV, we have summed in quadrature the errors from the $\Delta^*(1700)$ width and the branching ratio into $\gamma N$ given in the PDB \cite{Yao:2006px}. 
Table \ref{tab:raddecay} also shows the phases of the different contributions. The effective couplings from the $\Delta\pi$ and the $\Sigma^*K$ channels are almost in phase, i.e., they point in a similar direction, which makes them dominant over the $(\pi N)_d$ channel.

From the four different results in Tab. \ref{tab:raddecay} we prefer the decay width from fit 1. In the other fits some of the free parameters have been fixed. As argued in Sec. \ref{sec:improve}, the remaining free parameters of these extra fits have to absorb effects from this reduction of degrees of freedom, and results may become distorted. For fit 3 we have found a very similar solution, called fit 3', which has a reversed sign for the coupling of the $(\pi N)_d$-channel to the $\Delta^*(1700)$. This is a consequence of the weakness of this coupling as explained in Sec. \ref{sec:improve}. Despite the fact, that under the conditions of fit 1, we could not find such an alternative minimum in $\chi^2$, the two different decay widths of $459$ keV and $730$ keV from fit 3 and 3' give an idea of the intrinsic theoretical uncertainties of the present study. Thus, we assign a final value of 
\[
\Gamma=602\pm 140\; \mbox{keV}
\]
to the radiative decay width of the $\Delta^*(1700)$, with the error given by the difference between fit 3 and 3'.

The contribution from the $\Sigma^*K$ channel is smaller than from the $\Delta\pi$ channel as Tab. \ref{tab:raddecay} shows.
At first sight this seems surprising as $g_{\Sigma^*K}$ from Tab. \ref{tab:couplings} is four times larger than $g_{\Delta\pi}$. However, the threshold for this channel is at $\sqrt{s}=1880$ MeV; at $\sqrt{s}=1.7$ GeV it is closed and the $\Sigma^*K$ loops from Fig. \ref{fig:delstardec} are small. Obviously, the $\Delta\pi$ channel with a threshold much lower than the $\Delta^*(1700)$ mass dominates the decay; even $(N\pi)$ contributes despite to the weak coupling to the $\Delta^*(1700)$. A similar pattern has been observed in \cite{Doring:2006ub} for the radiative decay of the \lamm. As a consequence, the contribution from the $\rho N$ channel, which has not been considered in this study, will only moderately change the results.

In the present scheme, the large contributions from the $\Delta^{++}\pi^-$ and $\Delta^0\pi^+$ channel add up in the isospin combination from Eq. (\ref{firsttsdelstar}), and as a consequence the $\Delta\pi$ channel gives a large contribution to the radiative decay, in good agreement with experiment. This is in analogy to the decay $\Lambda^*(1520)\to\gamma\Sigma^0$ studied in \cite{Doring:2006ub} where the dominant $\pi \Sigma$ and $\pi\Sigma^*$ channels add up and result in good agreement with data. 

In Ref. \cite{Doring:2006ub}, it has also been observed that these channels cancel exactly for the $\Lambda^*(1520)\to\gamma\Lambda$ decay and the discrepancy between experiment and coupled channel model has been attributed to a genuine three quark component in the wave function of the \lamm, on top of the meson-baryon component from the coupled channel model. For the $\Delta^*(1700)$ there is no such additional channel for which we can test the model. However, there is a long history of calculating radiative decays of excited baryons in quark models, e.g. \cite{Koniuk:1979vy,Myhrer:1983eg,Capstick:2000qj,Merten:2002nz}. For the radiative decay of the $\Delta^*(1700)$ all these works obtain good agreement with experiment. 

Thus, the radiative decay appears well reproduced in both the quark model picture and the present scheme where the degrees of freedom are the mesons and baryons. A calculation of electroproduction within the present framework could help bring further insight into the question of which theoretical framework is more appropriate; e.g., the quark model from \cite{Merten:2002nz} has some difficulties in the $D_{33}$ channel, but improving the experimental data situation is certainly desirable.

\section{Conclusions}
In the study of the $\Delta^*(1700)$ radiative decay, a model has been formulated in which the photon couples to the final loops of the rescattering series that dynamically generates the $\Delta^*(1700)$. The $\pi N$ channel in $d$-wave has been included in the coupled channel scheme. Furthermore, the phototransitions for the dominant $s$-wave loops have been treated in a fully gauge invariant way.

Previous studies of eleven different pion- and photon-induced reactions have accumulated evidence of a strong coupling of the $\Delta^*(1700)$ to $\Delta\eta$ and $\Sigma^*K$, which is a prediction of the unitary coupled channel model. The present study provides an extra independent test for the nature of the $\Delta^*(1700)$, resulting in additional consistency for the picture in which the $\Delta^*(1700)$ is dynamically generated.
\subsection*{\bf Acknowledgments}
I would like to thank E. Oset for many suggestions and a critical reading of the paper.
This work is partly supported by DGICYT contract number
BFM2003-00856 and Generalitat Valenciana. This
research is  part of the EU Integrated Infrastructure Initiative 
Hadron Physics Project
under  contract number RII3-CT-2004-506078.
It has also been supported by the program {\it Formaci\'on de Profesorado Universitario} of the Spanish Government.

\end{document}